\documentclass[runningheads]{llncs}
\usepackage{graphicx}

\usepackage{url}
\usepackage[hidelinks]{hyperref}
\usepackage{fancyvrb}
\usepackage[final]{pdfpages}
\usepackage{times}
\usepackage{amsmath, amssymb}
\usepackage{multirow}
\usepackage{booktabs}
\usepackage{multirow}
\usepackage{graphicx}
\usepackage{subcaption}
\usepackage{float}
\usepackage{tabularx}
\usepackage{booktabs} 
\usepackage{orcidlink}
\renewcommand{\orcidID}[1]{\orcidlink{#1}}
\usepackage[T1]{fontenc}
\usepackage{graphicx}
\begin{document}
\title{Are Multimodal Embeddings Truly Beneficial for Recommendation? A Deep Dive into Whole vs. Individual Modalities}
%
%
\author{
  Yu Ye\inst{1}\orcidID{0009-0001-8556-6243} \and
  Junchen Fu\inst{1}\orcidID{0000-0003-4759-2042}\thanks{Corresponding author.} \and
  Yu Song\inst{2}\orcidID{0000-0002-8940-2561} \and
  Kaiwen Zheng\inst{1}\orcidID{0009-0007-2516-8407} \and
  Joemon M.~Jose\inst{1}\orcidID{0000-0001-9228-1759}
}

\authorrunning{Y.~Ye et al.}

\titlerunning{Are Multimodal Embeddings Truly Beneficial for Recommendation?}

\institute{
  University of Glasgow, Glasgow, United Kingdom\\
  \email{yu.jade.ye@gmail.com, j.fu.3@research.gla.ac.uk,%
  k.zheng.1@research.gla.ac.uk, joemon.jose@glasgow.ac.uk}
  \and
  Michigan State University, East Lansing, United States\\
  \email{songyu5@msu.edu}
}

\maketitle              
\begin{abstract}
Multimodal recommendation has emerged as a mainstream paradigm, typically leveraging text and visual embeddings extracted from pre-trained models such as Sentence-BERT, Vision Transformers, and ResNet. This approach is founded on the intuitive assumption that incorporating multimodal embeddings can enhance recommendation performance. However, despite its popularity, this assumption lacks comprehensive empirical verification. This presents a critical research gap. To address it, we pose the central research question of this paper: \textit{Are multimodal embeddings truly beneficial for recommendation?}

To answer this question, we conduct a large-scale empirical study examining the role of text and visual embeddings in modern multimodal recommendation models, both as a whole and individually.  Specifically, we pose two key research questions: \textit{(1) Do multimodal embeddings as a whole improve recommendation performance? (2) Is each individual modality—text and image—useful when used alone?}

To isolate the effect of individual modalities—text or visual—we employ a modality knockout strategy by setting the corresponding embeddings to either constant values or random noise. To ensure the scale and comprehensiveness of our study, we evaluate \textbf{14} widely used state-of-the-art multimodal recommendation models. Our findings reveal that: (1) multimodal embeddings generally enhance recommendation performance—particularly when integrated through more sophisticated graph-based fusion models. Surprisingly, commonly adopted baseline models with simple fusion schemes, such as VBPR and BM3, show only limited gains. (2) The text modality alone achieves performance comparable to the full multimodal setting in most cases, whereas the image modality alone does not. These results offer foundational insights and practical guidance for the multimodal recommendation community. Our code and datasets are available under \url{https://github.com/GAIR-Lab/MKF4MMRec}.

\end{abstract}

\section{Introduction}

Multimodal recommendation has emerged as a powerful paradigm in modern recommender systems, enabling the integration of diverse modalities such as text, image, video, and audio. By leveraging pretrained encoders such as Sentence-BERT, Vision Transformers (ViT), and ResNet, recent state-of-the-art models—including BM3~\cite{zhou2023bootstrap}, FREEDOM~\cite{zhou2023tale}, and SMORE~\cite{ong2025spectrum}—have demonstrated impressive performance across various recommendation tasks. These models are built on the intuitive yet widely accepted assumption that incorporating multimodal embeddings provides richer user-item representations and thus improves recommendation accuracy.

Despite the empirical success of these models, existing research has primarily concentrated on architectural innovations and fusion strategies~\cite{liu2024multimodal,zhou2023comprehensive,fu2025efficient,ge2024towards}, while largely neglecting a more fundamental issue—the contribution of multimodal embeddings themselves to the observed performance gains. In particular, a large number of multimodal recommendation models have introduced complex mechanisms such as graph-based modeling of modalities and intricate fusion architectures~\cite{zhuang2025bridging}. While these models are significantly more sophisticated than simple ID-based recommenders, if such complexity fails to yield tangible benefits, it raises a natural question: why use them at all? Although many works routinely integrate both text and image modalities, there is still a lack of comprehensive and controlled investigation into the actual utility of these embeddings. This presents a critical research gap. Motivated by this, we pose the central research question of this paper: \textit{Are multimodal embeddings truly beneficial for recommendation?}

Although extensive research has been conducted in the multimodal recommendation domain, evaluating the true effectiveness of multimodal embeddings remains challenging due to several factors~\cite{beel2016towards}: (1) most studies emphasize novel architectural designs or training strategies rather than directly examining the intrinsic quality of the embeddings; (2) evaluation protocols vary widely---including different datasets, metrics, and data-splitting strategies---making cross-study comparisons largely inconclusive; (3) reported results often reflect a combination of influences such as model architecture, optimization techniques, and training heuristics, which obscures the isolated contribution of the embeddings themselves.

To overcome the aforementioned limitations and answer the central question, we conduct a large-scale empirical study within a controlled and consistent experimental environment, focusing specifically on the role of multimodal embeddings in recommendation. We adopt a \textit{modality knockout} strategy, where the embedding of a given modality is replaced with either constant vectors (e.g., all zeros), or random noise. This simple yet effective approach allows us to isolate the contribution of each modality without introducing confounding architectural factors. Our investigation centers on two sub-questions: (1) Do multimodal embeddings as a whole enhance recommendation performance? and (2) Is each individual modality---text or image---beneficial when used in isolation? To address these questions, we apply our methodology to \textbf{14 widely-used multimodal recommendation models}, spanning a diverse set of fusion techniques and architectural families, ensuring a comprehensive and representative evaluation.

Our findings yield several notable insights. First, we confirm that leveraging multimodal embeddings generally enhances recommendation performance in most evaluated models equipped with sophisticated fusion mechanisms. However, we also observe some surprising results in methods employing simpler fusion strategies, such as VBPR and BM3, which are commonly used as baselines in multimodal recommendation. Second, we find that the text modality alone often performs comparably to the full multimodal setting, highlighting its dominant role. In contrast, the image modality alone provides only limited performance gains, and in many cases, its removal does not significantly degrade accuracy. The visualization of both text and image embeddings shows that text embeddings are more compact, whereas image embeddings are more dispersed, which may make them harder to learn.

In addition to these findings, we also summarize a set of practical techniques that help boost the performance of classic multimodal recommendation baselines, especially in resource-constrained or modality-limited scenarios.

Our contributions are threefold:
\begin{itemize}
    \item We present the first large-scale and systematic analysis of the role of multimodal embeddings in multimodal recommendation, evaluating 14 state-of-the-art models.
    \item We investigate the effectiveness of multimodal information from both whole and individual perspectives, providing empirical answers to long-standing assumptions in the field.
    \item We provide practical insights and a set of additional empirical findings that serve as actionable guidance for future multimodal recommendation model development and evaluation.
\end{itemize}

We will release all code and datasets to promote reproducibility and further exploration into the foundational assumptions of multimodal recommendation.

\section{Related Work}
\noindent \textbf{Reproducibility Study of Recommender Systems}  
In the field of recommender systems, differences in datasets, evaluation protocols, and experimental settings make reproducibility studies particularly important \cite{beel2016towards}. Rather than proposing new methodologies, these studies re-examine widely adopted baselines to draw fundamental insights for future research. For example, \cite{ferrari2019we} reveals that many state-of-the-art neural network-based recommendation methods perform even worse than simple nearest-neighbor techniques, \cite{rendle2020neural} revisits matrix factorization in comparison with neural collaborative filtering, and \cite{petrov2022systematic} conducts a reproducibility study on the BERT4Rec model~\cite{sun2019bert4rec}. \cite{ji2023critical} presents a critical study on the offline evaluation of recommender systems. In the recent modality-based recommendation domain, many works~\cite{li2023exploring,yuan2023go,fu2024exploring,ni2023content,zhang2024ninerec} compare advanced sequential recommendation models with heavy end-to-end encoders against simple ID embedding models, demonstrating that ID-based models still achieve strong performance. Although these studies do not propose novel methodologies, they provide valuable insights that have had a significant impact on the research community. Each serves as a cornerstone of the field, inspiring our study on recent state-of-the-art multimodal recommendation models.

\noindent \textbf{Multimodal recommendation}. 
Multimodal recommendation~\cite{liu2024multimodal,zhou2023comprehensive,fu2024iisan} leverages auxiliary information from different modalities, primarily text, and images, to enrich item representations. Early work like VBPR~\cite{he2016vbpr} extended the classic Bayesian Personalized Ranking (BPR~\cite{rendle2012bpr}) framework by incorporating pre-extracted visual features into the matrix factorization model, demonstrating that visual style can influence user preference. With the rise of Graph Neural Networks (GNNs), which excel at capturing high-order connectivity, graph-based models became the dominant paradigm. MMGCN~\cite{wei2019mmgcn} was a pioneering work that constructed separate user-item bipartite graphs for each modality, learning modality-specific representations through graph convolution. Following this, GRCN~\cite{wei2020graph} proposed refining the user-item graph by using multimodal features to identify and down-weight potential noisy (false-positive) interactions. More recent models like LATTICE~\cite{zhang2021mining} and FREEDOM~\cite{zhou2023tale} introduced an explicit item-item semantic graph built from multimodal features to capture latent item relationships, which is then combined with the user-item interaction graph. LGMRec~\cite{guo2024lgmrec} further extended this by learning both local and global user interests, using a hypergraph to model global dependencies. Other approaches, such as BM3~\cite{zhou2023bootstrap}, adopt a self-supervised learning paradigm. Its core innovation lies in bootstrapping latent representations by generating contrastive views through a simple dropout mechanism in the embedding space, thereby eliminating the need for explicit and computationally expensive graph augmentation or construction. TARec~\cite{zhuang2025bridging} and FDRec~\cite{zhuang2025frequency} adopt Wasserstein-based knowledge distillation for multimodal recommendation. IISAN~\cite{fu2024iisan,fu2025efficient} and CROSSAN~\cite{fu2025crossan} adopt a Decoupled PEFT structure for multimodal adaptation, matching full fine-tuning performance with reduced GPU memory usage.

\noindent \textbf{Concurrent work.}
Concurrent to ours, Pomo et al.~\cite{pomo2025mmrep} examine whether multimodal recommendation gains stem from semantics or capacity, introducing LVLM(Large Vision–Language Model)-derived, multimodal-by-design embeddings via structured VQA prompting with EOS-based extraction.
Meanwhile, Zhou et al.~\cite{zhou2025does} propose an evaluation framework and benchmark multimodal recommendation models across tasks, stages, and integration strategies to assess modality contributions.
By contrast, our work diagnoses reliance via \emph{Modality Knockout}: we replace the target modality with constants/noise during both training and inference while preserving the original model architecture, isolating data-signal effects and enabling consistent, model-agnostic comparison.


Despite increasing architectural complexity, many methods implicitly assume that multimodal embeddings improve recommendation performance, while rigorous tests of this assumption remain limited. To our knowledge, ours is a pioneering effort in large-scale, systematic investigations that isolate modality information via architecture-preserving knockouts, providing clear evidence of when—and to what extent—text and image embeddings matter.

\section{Preliminaries}
\label{sec:preliminaries}
\noindent \textbf{Problem Definition.} This paper investigates the role of multimodal embeddings in multimodal recommendation. Let \(U\) denote the set of users and \(I\) the set of items. In the multimodal recommendation setting, each item \(i \in I\) is associated with both visual and textual embeddings. Specifically:
\begin{itemize}
  \item \textbf{Visual Modality:} a visual embedding \(\mathbf{e}_{v_i} \in \mathbb{R}^d\), extracted from item images using a pre-trained convolutional neural network (e.g., ResNet50) or a vision transformer.
  \item \textbf{Textual Modality:} a textual embedding \(\mathbf{e}_{t_i} \in \mathbb{R}^d\), extracted from item descriptions or titles using a model such as Sentence-BERT.
\end{itemize}
A multimodal recommendation model enhances its predictive performance by incorporating both visual and textual embeddings into the representation learning process. We denote the multimodal embedding by:
\[
  \tilde{\mathbf{e}}_{m_i}
  = (\mathbf{e}_{v_i},\,\mathbf{e}_{t_i}),
\]
where \((\cdot,\cdot)\) indicates a combination of the visual and textual embeddings, rather than a dimensional concatenation. The combined embedding \(\tilde{\mathbf{e}}_{m_i}\) is then fed into a recommendation model, which may integrate it with user and item ID embeddings through various strategies such as simple fusion (e.g., VBPR~\cite{he2016vbpr}), incorporation into a user–item interaction graph (e.g., MMGCN~\cite{wei2019mmgcn}), or by constructing an item–item similarity graph (e.g., FREEDOM~\cite{zhou2023tale}). In this work, we dissect the contribution of each modality by analyzing their impact from two perspectives:
\begin{enumerate}
  \item \textbf{Overall Contribution:} We evaluate the importance of the multimodal embedding \(\tilde{\mathbf{e}}_{m_i}\) as a whole.
  \item \textbf{Individual Contribution:} We quantify the individual effects of the visual embedding \(\mathbf{e}_{v_i}\) and the textual embedding \(\mathbf{e}_{t_i}\) individually.
\end{enumerate}

\noindent \textbf{Modality Knockout.} To rigorously answer our research question, we adopt a simple yet effective diagnostic methodology we named ``Modality Knockout''. The central idea is to systematically replace the information from specific modalities and measure the resulting impact on the model's final performance. Our method is inspired by \cite{he2022masked}, which examines scenarios where only a small subset of image patches is utilized to assess their representational capability, while the remaining patches are replaced with random noise.

We implement this by substituting the original, informative modality embeddings with arbitrary vectors—either Gaussian noise or constant vectors (e.g., all zeros or all ones)—during both training and inference. This approach creates controlled experimental conditions to evaluate the model’s reliance on each modality. We consider three primary “knockout” scenarios:

\begin{enumerate}
    \item \textbf{Visual Knockout:} Replace only the visual embeddings, \(e_{v_i}\), preserving textual and collaborative signals, to gauge performance without visual information.
    \item \textbf{Textual Knockout:} Replace only the textual embeddings, \(e_{t_i}\), retaining visual and collaborative signals, to measure performance in the absence of text.
    \item \textbf{Dual Knockout:} Replace both visual and textual embeddings, \(\tilde{\mathbf{e}}_{m_i}\), isolating the contribution of the collaborative (ID) embeddings alone.
\end{enumerate}

\section{Experimental Setup}

\subsection{Datasets}
\begin{table}[t!]
\centering
\caption{Statistics of the Datasets.}
\begin{tabular}{lcccc}
\toprule
\multirow[c]{2}{*}{\textbf{Datasets}} & \multirow[c]{2}{*}{\textbf{$|User|$}} & \multirow[c]{2}{*}{\textbf{$|Item|$}} & \multirow[c]{2}{*}{\textbf{$|Inter|$}} & \multirow[c]{2}{*}{\textbf{Sparsity (\%)}} \\
\\
\midrule
Baby     & 19445  & 7050   & 160792  & 99.88\% \\
Sports   & 35598  & 18357  & 296337  & 99.95\% \\
Clothing & 39387  & 23033  & 278677  & 99.97\% \\
\bottomrule
\end{tabular}
\end{table}

We conduct our experiments on three widely-used Amazon~\cite{ni2019justifying} datasets, which are standard benchmarks in multimodal recommendation research, including  \textbf{Baby}, \textbf{Clothing}, and \textbf{Sports}. Each dataset provides user--item interactions, item metadata including textual descriptions, and item images. We follow the standard data preprocessing steps used in the MMRec library\footnote{\url{https://github.com/enoche/MMRec/}} to ensure fair comparison, including 5-core filtering and splitting data into training, validation, and test sets. 

Item image features were obtained by using the pre-extracted vectors extracted by ResNet-50~\cite{he2016deep} provided in the benchmark; these serve as the baseline visual embeddings. Item text features typically come from product titles or descriptions, encoded via pre-trained all-MiniLM-L6-v2 model~\cite{reimers-2020-sentence-bert}.

\subsection{Models Implemented via the MMRec Toolbox}

We selected a representative set of influential and high-performing multimodal recommendation models, encompassing diverse architectural philosophies. Our focus is on how these models handle multimodal embeddings in recommendation, which can be broadly categorized into four paradigms:  

\begin{enumerate}
    \item \textbf{Simple fusion approaches (Group 1):} Traditional approaches integrate multimodal data in a straightforward manner—without graph-based fusion—by combining ID embeddings with other modalities via simple concatenation or direct contrastive learning. This group includes \textbf{VBPR}~\cite{he2016vbpr} and \textbf{BM3}~\cite{zhou2023bootstrap}.
    
    \item \textbf{User--item graph integration (Group 2):} A representative work in this category is the popular \textbf{MMGCN}~\cite{wei2019mmgcn}, which constructs separate user--item graphs for each modality to leverage multimodal signals. This group includes \textbf{GRCN}~\cite{wei2020graph}, \textbf{DualGNN}~\cite{wang2021dualgnn}, \textbf{SLMRec}~\cite{tao2022self}, \textbf{MMGCN}~\cite{wei2019mmgcn}, and \textbf{LGMRec}~\cite{guo2024lgmrec}.
    
    \item \textbf{Item--item graph construction (Group 3):} Another paradigm exploits multimodal embeddings to build item-item graphs, capturing relationships between items through their multimodal content. This group includes \textbf{LATTICE}~\cite{zhang2021mining}, \textbf{FREEDOM}~\cite{zhou2023tale}, \textbf{MGCN}~\cite{yu2023multi}, \textbf{DA-MRS}~\cite{xv2024improving}, and \textbf{SMORE}~\cite{ong2025spectrum}.
    
    \item \textbf{Hybrid graph approaches (Group 4):} Another line of research combines the advantages of the above strategies, incorporating multimodal information into both user-item and item-item graph construction, thereby enhancing representation learning from both perspectives. This group includes \textbf{DRAGON}~\cite{zhou2023enhancing} and \textbf{PGL}~\cite{yu2025mind}.
\end{enumerate}

\subsection{Experimental Protocol}

We adopt the Model implementation, data split with the pre-extracted multimodal embeddings, evaluation implementation, and hyperparameter tuning strategies, from MMRec\footnote{\url{https://github.com/enoche/MMRec/}} to ensure a fair comparison for all models. 

To investigate the role of multimodal embeddings, we employ Modality Knockout as introduced in \autoref{sec:preliminaries} and define four experimental conditions. For each knockout setting, we consider both constant-value and random-noise knockouts.\footnote{For constant-value knockouts, we primarily use ones as the default, since using zeros as the default can be incompatible with certain model implementations and may cause training to fail.
} Specifically, we include these four settings in the experiments:
\begin{enumerate}
    \item \textbf{Baseline (V+T)}: The model is trained and evaluated normally, using both visual (V) and textual (T) features. This represents the performance reported in the original papers.

    \item \textbf{Visual Knockout (T-Only)}: The model is trained and evaluated with the visual modality completely degraded. We achieve this by replacing all image feature embeddings with a vector of constant values or random noise. The text modality remains unchanged. This setup tests the model's performance when it can only rely on textual and collaborative information.

    \item \textbf{Textual Knockout (V-Only)}: Similarly, we degrade the textual modality by replacing all text feature embeddings, while keeping the visual modality intact. This tests the model's reliance on visual and collaborative signals.

    \item \textbf{Dual Knockout (ID-Only)}: Both visual and textual embeddings are replaced. This isolates the pure collaborative filtering signal derived from the user-item interaction structure.
\end{enumerate}

All models are trained to converge on the training set, and performance is reported on the test set using standard top-K recommendation metrics \cite{he2017neural} including \textbf{Recall} and \textbf{NDCG}, and \textbf{Precision}.

\begin{figure*}[t]
  \centering
  \includegraphics[width=\textwidth]{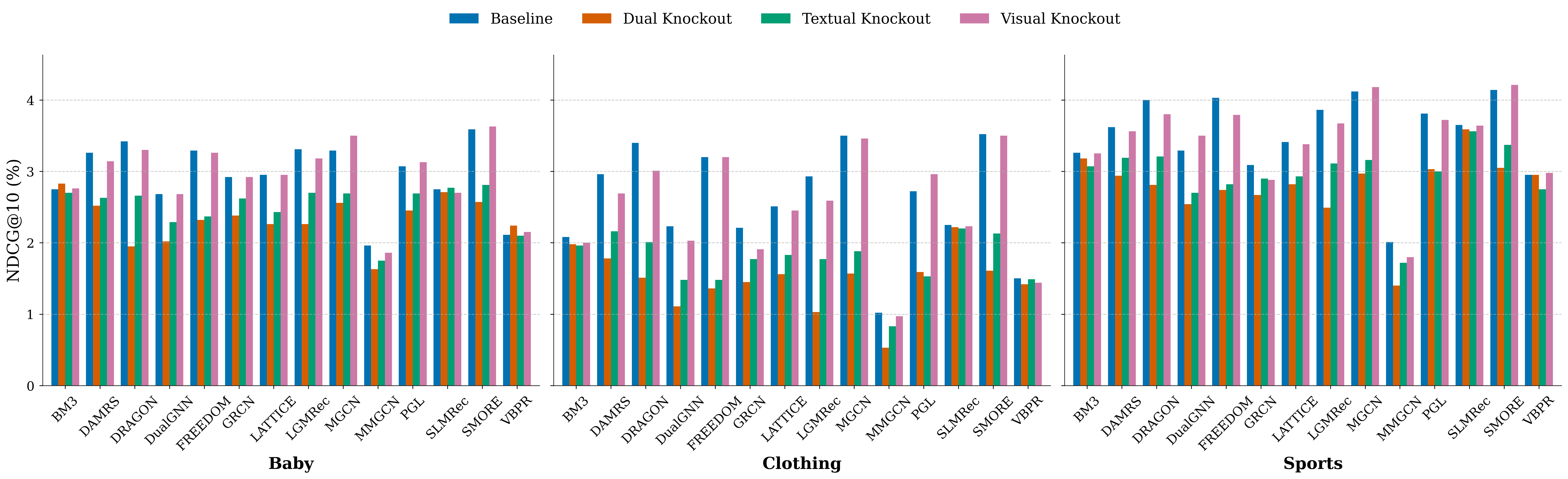}
  \caption{Comparing 14 Multimodal Recommendation Models under Modality Knockouts on Baby, Clothing, and Sports Datasets.}
  \label{fig:ndcg10_all_rq1_rq2}
\end{figure*}

\begin{figure*}[t]
  \centering
  \includegraphics[width=\textwidth]{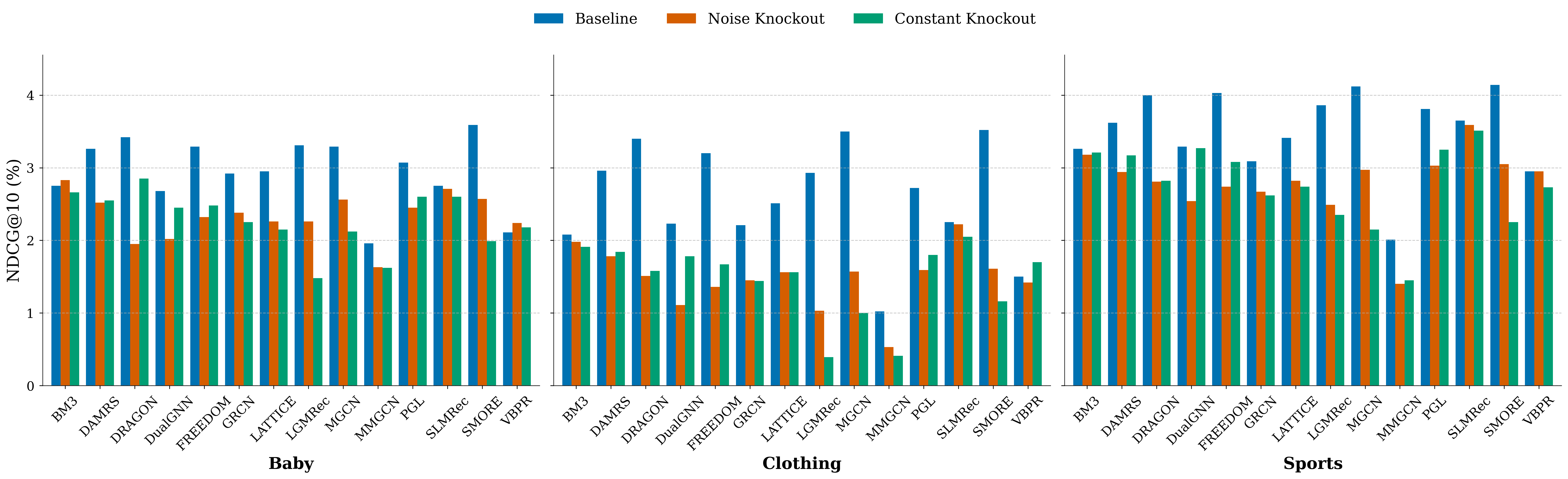}
  \caption{Comparing Noise-Based and Constant Embedding Replacement with Baseline with Dual Knockout. }
  \label{fig:dual_knockout_types}
\end{figure*}

\section{Main Experiment Results}

In this section, we present the results of our Modality Knockout experiments to answer two primary research questions (RQs). We analyze the overall utility of multimodal information and then dissect the individual contributions of the visual and textual modalities.

\begin{description}
  \item[\textbf{RQ1:}] \textit{Do multimodal embeddings as a whole improve recommendation performance?}

  \item[\textbf{RQ2:}] \textit{Is each individual modality—text and image—useful when used alone?}

\end{description}

\subsection{Multimodality as a Whole (RQ1)}
\label{sec:rq1}

To answer this question, we analyze the \textbf{Dual Knockout} scenario, where both visual and textual embeddings are replaced with uninformative vectors. This allows us to measure the total contribution of multimodal information by observing the performance drop when it is entirely removed, leaving only the collaborative signal from user/item IDs.

As shown in the ``Dual Knockout'' experiment results in \autoref{fig:ndcg10_all_rq1_rq2} and \autoref{fig:dual_knockout_types}, nearly all models across the three datasets experience substantial performance degradation on both knockout types with constant values and random-noise vectors, with the exception of BM3, SLMRec, and VBPR. Notably, both VBPR and BM3 belong to Group~1, which adopts a simple fusion approach, whereas most models from the other groups exhibit clear performance drops.
This phenomenon will be discussed in detail in \autoref{sec:vbpr}. For instance, on the \textit{Baby} dataset, models like \textit{FREEDOM}, \textit{DRAGON}, and \textit{LGMRec} see their \textbf{NDCG@10} scores plummet by \textbf{29.5\%}, \textbf{43.0\%}, and \textbf{31.7\%}, respectively. Similar trends are observed on the \textit{Clothing} and \textit{Sports} datasets, with performance drops typically ranging from \textbf{20\% to over 60\%}.

\textbf{(Answer to RQ1)} This demonstrates that the joint use of visual and textual features as a whole provides crucial information that substantially improves recommendation performance in most multimodal recommendation models with sophisticated fusion. Interestingly, common baselines with simpler fusion strategies, such as VBPR and BM3, exhibit surprisingly "stable" performance, suggesting that the incorporated modality information is not beneficial in these models.

\begin{figure*}[t]
  \centering
    
  \begin{subfigure}{0.32\textwidth}
    \centering
    \includegraphics[width=\linewidth]{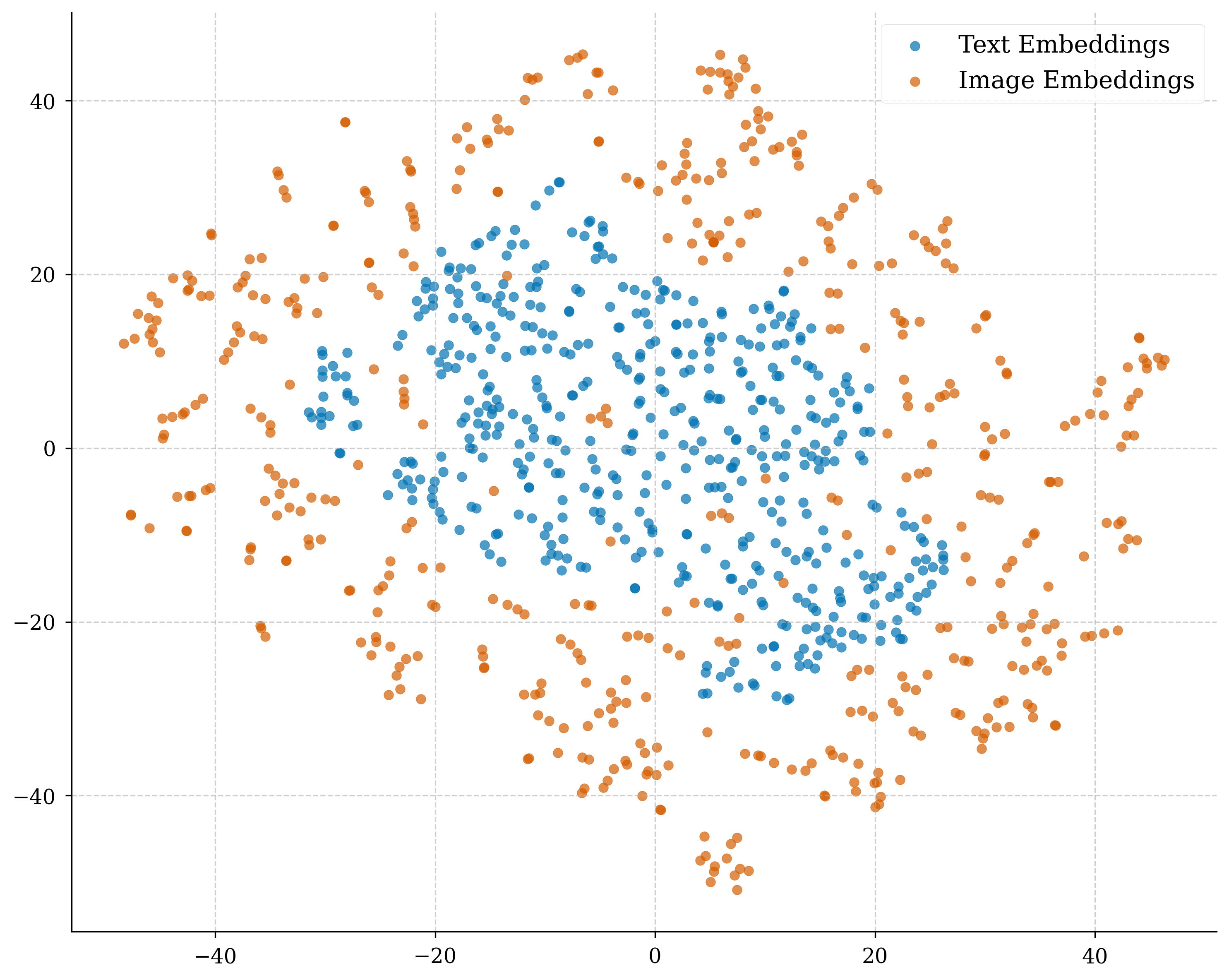}
    \caption{Baby}
    \label{fig:sub1}
  \end{subfigure}
  \hfill
  \begin{subfigure}{0.32\textwidth}
    \centering
    \includegraphics[width=\linewidth]{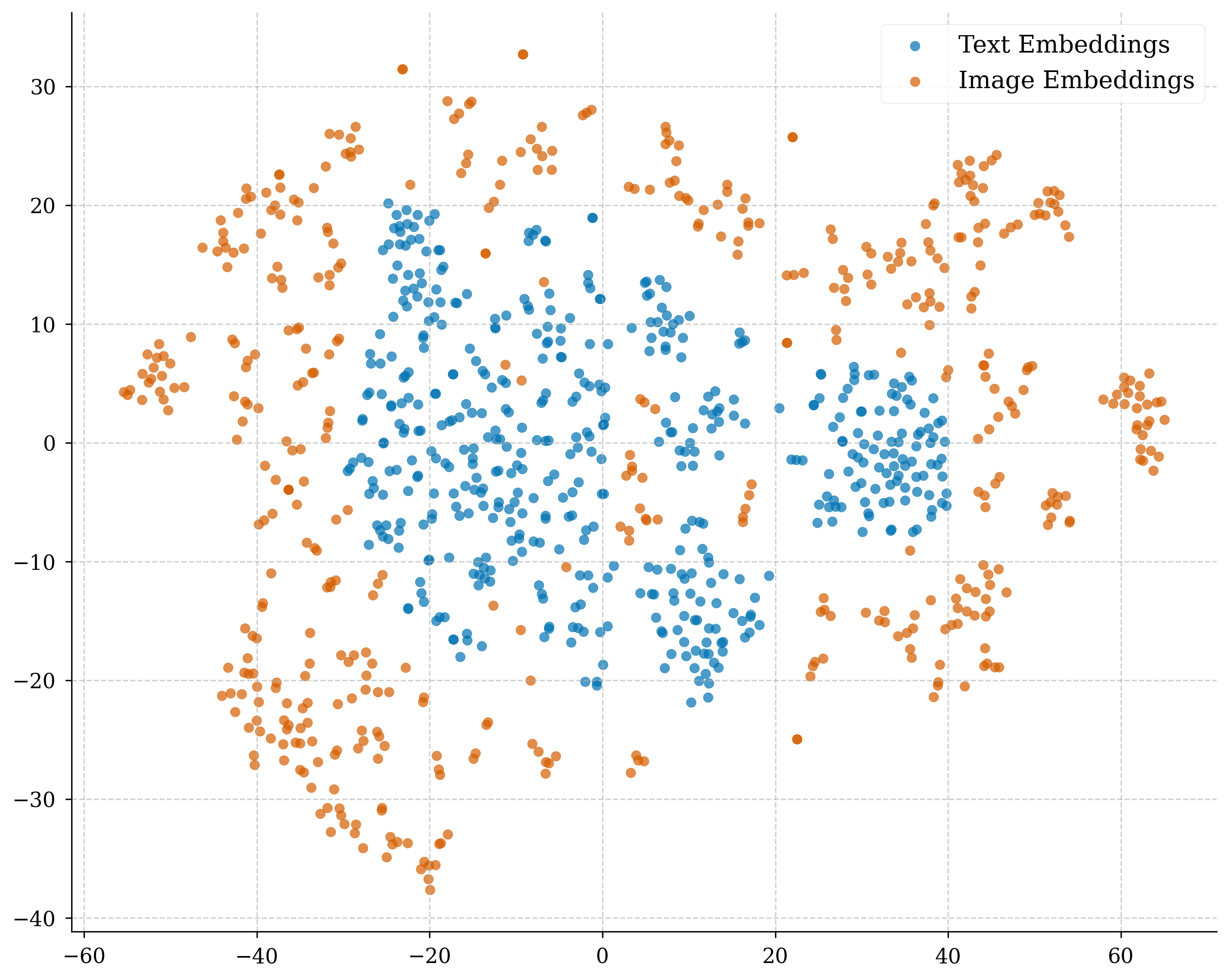}
    \caption{Clothing}
    \label{fig:sub2}
  \end{subfigure}
  \hfill
  \begin{subfigure}{0.32\textwidth}
    \centering
    \includegraphics[width=\linewidth]{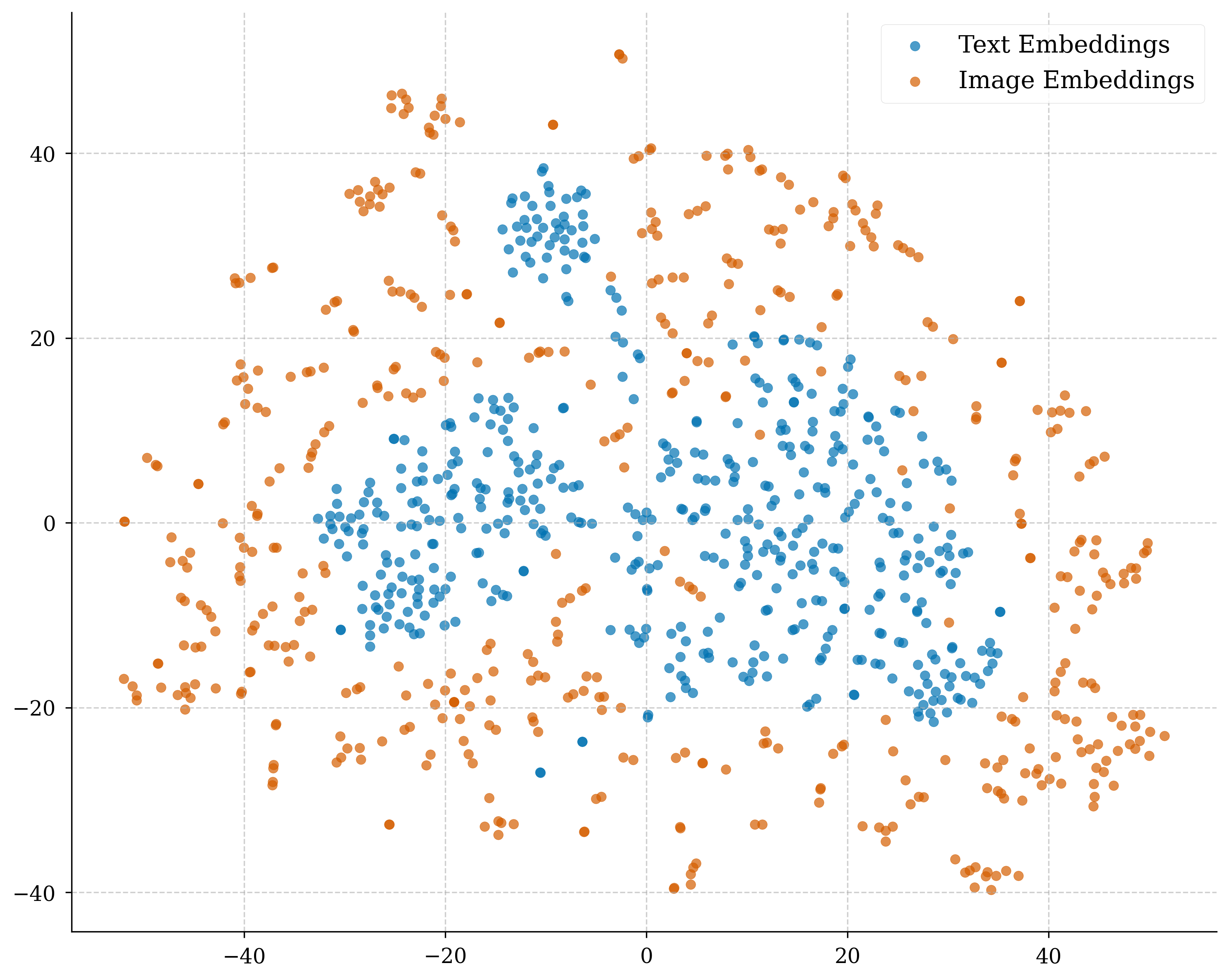}
    \caption{Sports}
    \label{fig:sub3}
  \end{subfigure}
  \caption{Visualization of Text and Image Embeddings.}
  \label{fig:embedding_visualization}
\end{figure*}

\subsection{Individual Modalities (RQ2)}

Having established the overall utility of multimodality, we now investigate the individual roles of the visual and textual modalities. We compare the results from the \textbf{Visual Knockout} and \textbf{Textual Knockout} experiments. The results reveal a striking asymmetry in how models depend on these two modalities. As demonstrated in \autoref{sec:rq1}, the two knockout strategies exhibit similar trends. Due to space constraints, we present only the results for the random noise–based knockout. 

\textbf{The Dominance of Text:} The \textit{Textual Knockout} experiment, where only text embeddings were replaced, triggered a severe performance drop across almost all models, comparable to the drop seen in the \textit{Dual Knockout} experiment. For example, on the \textit{Baby} dataset, \textit{FREEDOM}'s \textbf{NDCG@10} dropped by \textbf{28.0\%}, \textit{DRAGON}'s fell by \textbf{22.2\%} and \textit{LGMRec}'s fell by \textbf{18.4\%}. This clearly indicates that models are heavily reliant on the textual modality. The rich semantic information contained in item titles and descriptions appears to be the primary driver of performance gains in most modern architectures.

\textbf{The Ineffectiveness of Vision:} In stark contrast, the \textit{Visual Knockout} experiment, where only image embeddings were replaced, had a minimal effect on the majority of models. In some cases, such as \textit{PGL} on the \textit{Clothing} dataset, removing the visual information even led to a slight performance \textit{increase}. This strongly suggests that for a large portion of state-of-the-art models, the visual modality is severely underutilized, contributing little to the final recommendation. The models are not effectively ``seeing'' the images.

To analyze the differences between the two modalities, we visualize their distributions in \autoref{fig:embedding_visualization}. Both embeddings are normalized to address the mismatch caused by pre-extracted text embeddings being normalized while image embeddings are not. The figures reveal that text embeddings form a more compact distribution, whereas image embeddings are more dispersed, which may make the learning process more challenging.

\textbf{(Answer to RQ2)} Although the models are integrating multimodal embeddings, they operate in a state of textual dominance. The textual signal is easier to learn and consequently overshadows the visual signal during training, leaving the latter largely underutilized.

\section{Discussion}

\subsection{Visually-Sensitive Models}
In the RQ2 analysis, we observed that most models make limited use of visual-modality information. Nevertheless, we also identified a subset of models from Group 2 that are highly sensitive to the image modality—when their image embeddings were replaced, their performance degraded substantially. Based on the experimental results, we summarize:

\begin{table*}[t!]
\centering
\caption{Performance Comparison under Image Knockout for Vision-Sensitive Models.}
\label{tab:image_knockout_sensitive}
\begin{tabular}{llccc}
\toprule
\multirow{2}{*}{\textbf{Model}} & \multirow{2}{*}{\textbf{Dataset}} & \textbf{NDCG@10} & \textbf{Precision@10} & \textbf{Recall@10} \\
 & & Baseline / Knockout & Baseline / Knockout & Baseline / Knockout \\
\midrule
\multirow{3}{*}{GRCN} 
  & Baby     & 0.0292 / 0.0292 (+0.0\%) & 0.0059 / 0.0060 (+1.7\%) & 0.0533 / 0.0545 (+2.3\%) \\
  & Clothing & 0.0221 / 0.0191 (-13.6\%) & 0.0043 / 0.0038 (-11.6\%) & 0.0417 / 0.0364 (-12.7\%) \\
  & Sports   & 0.0309 / 0.0288 (-6.8\%)  & 0.0062 / 0.0058 (-6.5\%)  & 0.0557 / 0.0525 (-5.7\%) \\
\midrule
\multirow{3}{*}{LGMRec} 
  & Baby     & 0.0331 / 0.0318 (-3.9\%)  & 0.0065 / 0.0065 (+0.0\%)  & 0.0595 / 0.0592 (-0.5\%) \\
  & Clothing & 0.0293 / 0.0259 (-11.6\%) & 0.0055 / 0.0050 (-9.1\%)  & 0.0535 / 0.0482 (-9.9\%) \\
  & Sports   & 0.0386 / 0.0367 (-4.9\%)  & 0.0078 / 0.0074 (-5.1\%)  & 0.0710 / 0.0678 (-4.5\%) \\
\midrule
\multirow{3}{*}{MMGCN} 
  & Baby     & 0.0196 / 0.0186 (-5.1\%)  & 0.0041 / 0.0039 (-4.9\%)  & 0.0365 / 0.0350 (-4.1\%) \\
  & Clothing & 0.0102 / 0.0097 (-4.9\%)  & 0.0021 / 0.0020 (-4.8\%)  & 0.0202 / 0.0191 (-5.4\%) \\
  & Sports   & 0.0201 / 0.0180 (-10.4\%) & 0.0042 / 0.0038 (-9.5\%)  & 0.0370 / 0.0335 (-9.5\%) \\
\bottomrule
\end{tabular}
\end{table*}

\begin{itemize}
  \item \textbf{GRCN}: This model (Graph-Refined Convolutional Network) showed a clear dependency on images. On \textit{Clothing}, its NDCG@10 dropped about $-13.6\%$ from real images to noise. On \textit{Sports}, a smaller drop ($\sim-6.8\%$) was observed. These are substantial differences, indicating GRCN was indeed leveraging visual features to refine its recommendations. GRCN’s design of dynamically modifying the graph likely made it sensitive to the additional signal images provided (perhaps in deciding which interactions to prune or keep).

  \item \textbf{LGMRec}: The Local-Global Graph Learning model was among the strongest exploitation of images. For \textit{Clothing}, we saw roughly a $-11.6\%$ hit in NDCG@10. \textit{Sports} and \textit{Baby} had around $-3.9\%$ and $-4.9\%$ respectively. LGMRec's separate handling of modality-specific embeddings seems to pay off; when those embeddings carry no useful information, the model's predictive accuracy falls noticeably.

  \item \textbf{MMGCN}: The classic multi-modal GCN also showed reliance on images, though slightly less than the above two. Its NDCG@10 fell by about $4.9\%$--$10.4\%$ without images across the datasets. This is still a meaningful drop, confirming that MMGCN’s approach of modeling user preference per modality was indeed capturing something useful from images.
\end{itemize}

We believe this drop stems not just from the presence of image features, but more fundamentally from the models' architectural reliance on them, especially the fusion strategies they adopt for visual information.

Our results indicate that the three Group 2 models (GRCN, LGMRec, and MMGCN) exhibit distinct behavior compared to other models: they are highly sensitive to visual features. In practice, this means that deploying these models inherently places greater emphasis on visual information, making them particularly suitable for scenarios with abundant and reliable image data. By contrast, models in other groups show weaker dependence on visual signals, which makes them more resilient when visual information is sparse or noisy.

\subsection{Impact of Visual Encoder Choice: Replacing ResNet-50 with ViT}
To examine whether our conclusions are specific to the ResNet-50 visual features, we replace the image embeddings with ViT-based representations while keeping the datasets, splits, hyperparameters, and textual embeddings unchanged. Our results show that switching to ViT does \emph{not} yield uniform gains; instead, the effect is highly model-dependent. For several models, performance changes are marginal, suggesting limited sensitivity to the visual encoder. For example, BM3’s NDCG@10 changes by only +0.7\% / +0.5\% / $-1.2$\% on Baby/Clothing/Sports, and SMORE remains similarly stable (+1.1\% / $-0.9$\% / $-0.2$\%). In contrast, a subset of models exhibits strong encoder sensitivity, with either consistent improvements or notable degradations. GRCN provides a clear positive case, improving NDCG@10 by +1.7\% / +3.2\% / +2.9\% and Recall@10 +1.5\% / +2.6\% / +3.6\% on Baby/Clothing/Sports, indicating that richer ViT features can translate into better ranking accuracy when the architecture effectively exploits visual signals. Meanwhile, several models degrade substantially under ViT on specific datasets. DRAGON drops by $-11.1$\% in NDCG@10 on Baby and $-5.2$\% on Sports, MMGCN shows a pronounced decline on Clothing ($-8.8$\% NDCG@10 and $-9.4$\% Recall@10), and LGMRec decreases by $-7.8$\% on Clothing and $-4.1$\% on Sports. This mixed pattern suggests that the benefit of stronger visual representations is not only a function of feature quality, but also depends on how each model integrates visual information, which may require encoder-specific normalization or retuning to remain effective.

Overall, Our results reinforce that visual modality utility is architecture and mechanism dependent: some models can consistently convert stronger visual embeddings into gains, while others remain insensitive or even negatively impacted when the visual feature space changes.

\subsection{Case Study: VBPR’s Failure to Utilize Modality Features}
\label{sec:vbpr}

\begin{table*}[t!]
\centering
\caption{Performance Impact of Modality Knockouts on BM3, SLMRec, and VBPR.}
\label{tab:dualknockout_discussion}
\resizebox{\textwidth}{!}{
\begin{tabular}{ll|cccc|cccc|cccc}
\toprule
\multirow{2}{*}{\textbf{Model}} & \multirow{2}{*}{\textbf{Dataset}} & \multicolumn{4}{c|}{\textbf{NDCG@10}} & \multicolumn{4}{c|}{\textbf{Precision@10}} & \multicolumn{4}{c}{\textbf{Recall@10}} \\
& & Baseline & Text & Image & Dual & Baseline & Text & Image & Dual & Baseline & Text & Image & Dual \\
\midrule
\multirow{3}{*}{\centering BM3} 
& Baby     & 0.0275 & 0.0270 & 0.0276 & 0.0283 & 0.0057 & 0.0056 & 0.0057 & 0.0059 & 0.0517 & 0.0506 & 0.0516 & 0.0533 \\
& Clothing & 0.0208 & 0.0196 & 0.0200 & 0.0198 & 0.0039 & 0.0038 & 0.0039 & 0.0039 & 0.0375 & 0.0368 & 0.0369 & 0.0371 \\
& Sports   & 0.0326 & 0.0307 & 0.0325 & 0.0318 & 0.0066 & 0.0062 & 0.0064 & 0.0063 & 0.0603 & 0.0566 & 0.0587 & 0.0575 \\
\midrule
\multirow{3}{*}{\centering SLMRec}
& Baby     & 0.0275 & 0.0277 & 0.0270 & 0.0271 & 0.0054 & 0.0055 & 0.0053 & 0.0054 & 0.0492 & 0.0499 & 0.0483 & 0.0485 \\
& Clothing & 0.0225 & 0.0220 & 0.0223 & 0.0222 & 0.0044 & 0.0042 & 0.0043 & 0.0043 & 0.0422 & 0.0406 & 0.0416 & 0.0409 \\
& Sports   & 0.0365 & 0.0356 & 0.0364 & 0.0359 & 0.0073 & 0.0071 & 0.0072 & 0.0071 & 0.0663 & 0.0640 & 0.0652 & 0.0642 \\
\midrule
\multirow{3}{*}{\centering VBPR}
& Baby     & 0.0211 & 0.0210 & 0.0222 & 0.0224 & 0.0045 & 0.0045 & 0.0047 & 0.0047 & 0.0402 & 0.0399 & 0.0418 & 0.0417 \\
& Clothing & 0.0150 & 0.0149 & 0.0144 & 0.0142 & 0.0029 & 0.0029 & 0.0028 & 0.0028 & 0.0278 & 0.0271 & 0.0266 & 0.0261 \\
& Sports   & 0.0295 & 0.0275 & 0.0298 & 0.0295 & 0.0061 & 0.0056 & 0.0061 & 0.0060 & 0.0547 & 0.0503 & 0.0546 & 0.0543 \\
\bottomrule
\end{tabular}
}
\end{table*}

Furthermore, experiments on RQ1 revealed an intriguing pattern across several models, including BM3, SLMRec, and VBPR. When both textual and visual modalities were simultaneously disabled using the Dual Knockout approach, the models’ performance showed no substantial decline. Instead, the results fluctuated only slightly around the baseline, suggesting that these models do not effectively leverage the available multimodal information. Consequently, their predictive power appears to rely predominantly on user–item interaction data.

\begin{figure*}[t]
  \centering
  \begin{subfigure}{0.32\textwidth}
    \centering
    \includegraphics[width=\linewidth]{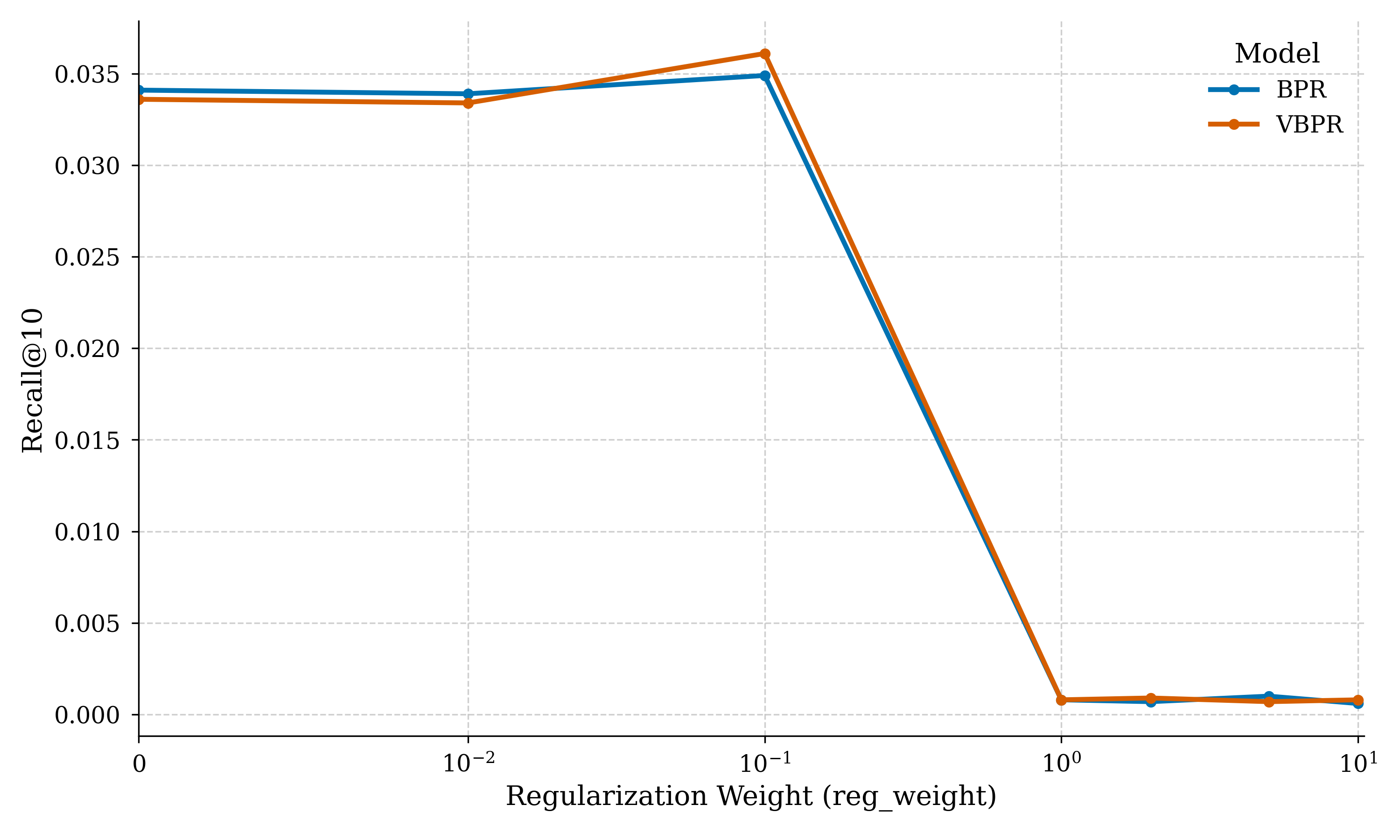}
    \caption{BPR vs. VBPR with Dual Knockout.}
    \label{fig:sub1}
  \end{subfigure}
  \hfill
  \begin{subfigure}{0.32\textwidth}
    \centering
    \includegraphics[width=\linewidth]{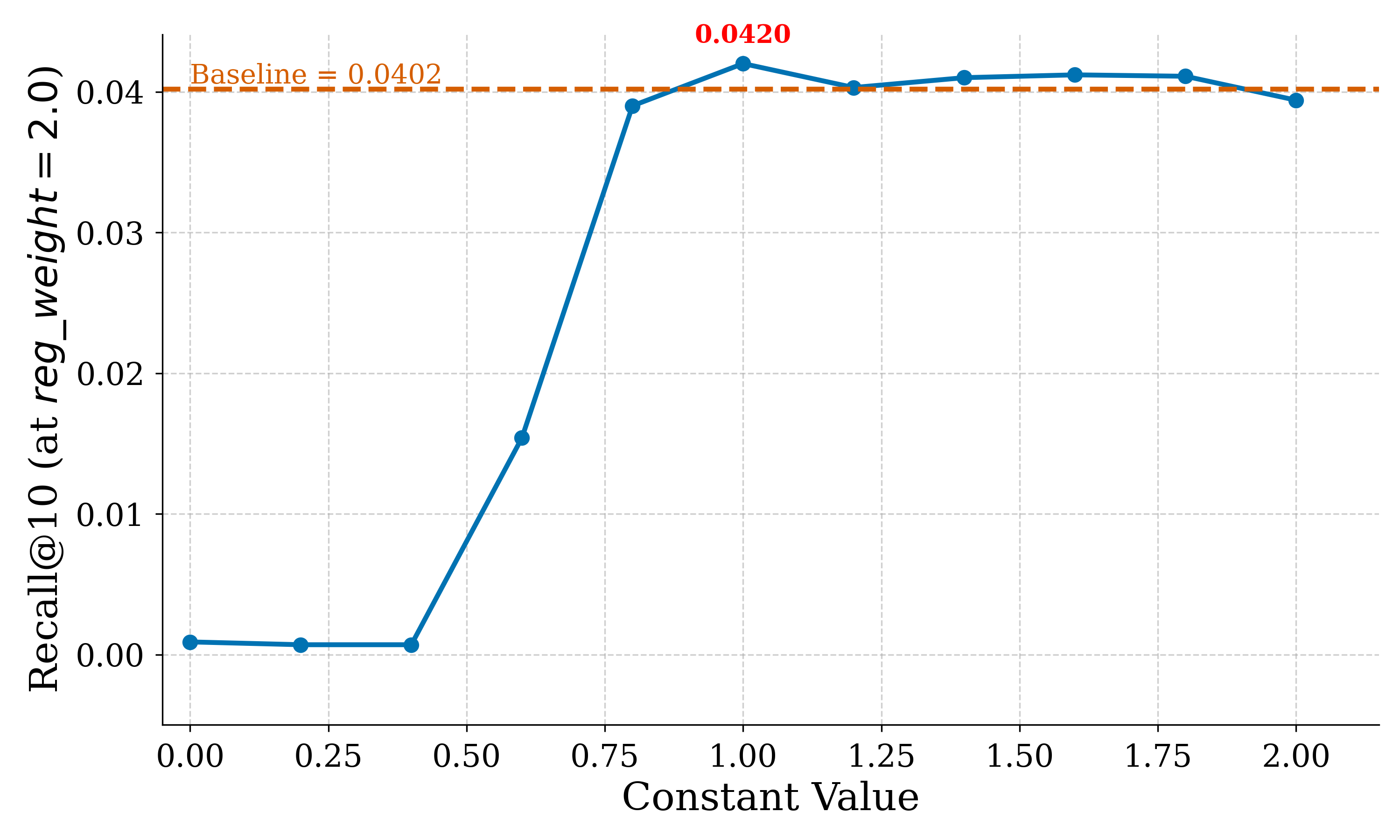}
    \caption{VBPR with Constant Embedding.}
    \label{fig:sub2}
  \end{subfigure}
  \hfill
  \begin{subfigure}{0.32\textwidth}
    \centering
    \includegraphics[width=\linewidth]{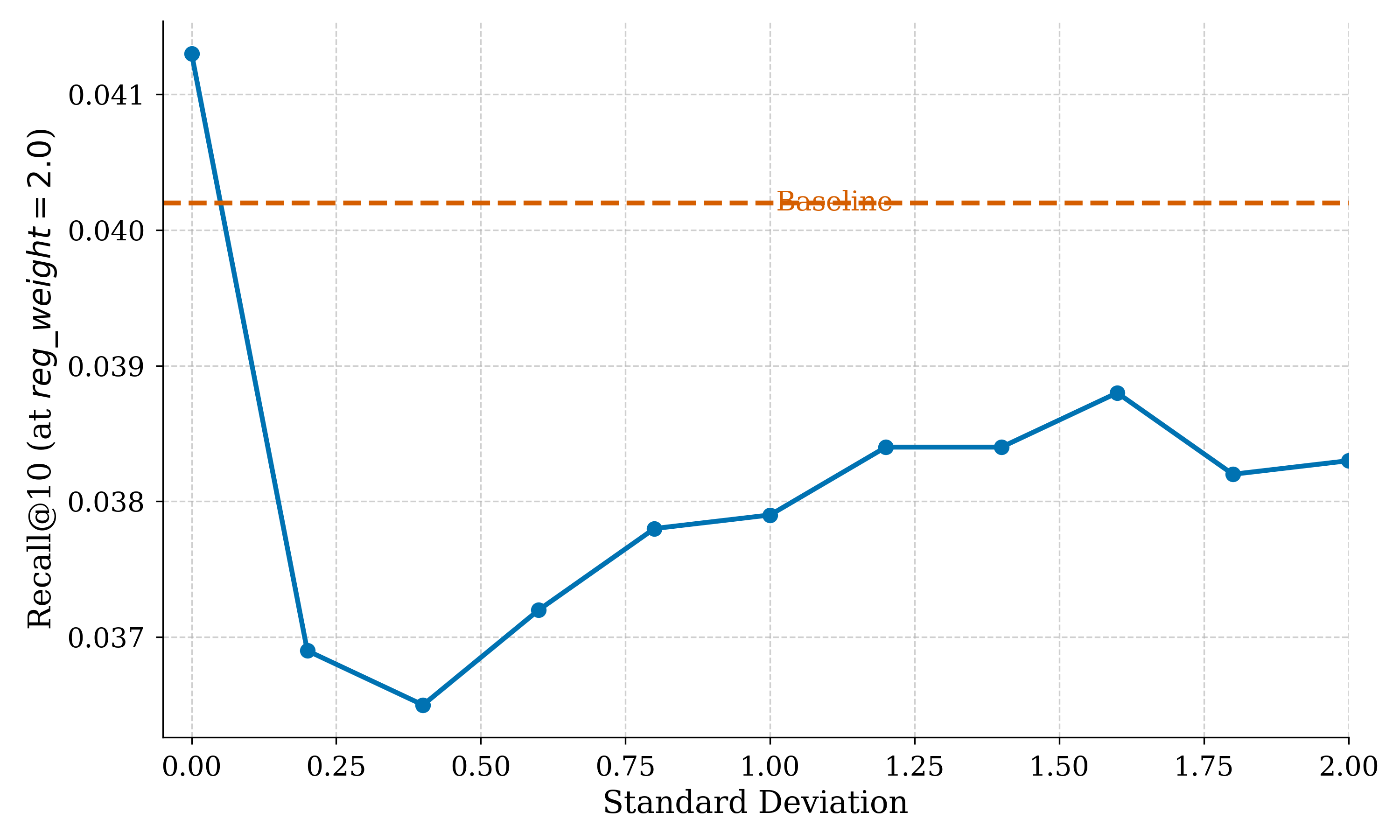}
    \caption{VBPR with Noise Embedding.}
    \label{fig:sub3}
  \end{subfigure}
  \caption{Recall@10 under Different Embedding Replacement Strategies.}
  \label{fig:threeplots}
  \vspace{-0.1in}
\end{figure*}

For instance, as described in \autoref{tab:dualknockout_discussion} under Dual Knockout, where features were substituted with Gaussian
noise, the changes in Recall@10 were surprisingly minimal. On the Clothing dataset, BM3's
performance dropped by only 1.1\%, while on Sports, it decreased by 4.6\%. Remarkably, on the
Baby dataset, its performance slightly increased by 3.1\%. Similarly, SLMRec's performance
changed by -3.1\% (Clothing), -3.2\% (Sports), and -1.4\% (Baby). VBPR showed changes of -6.1\%
(Clothing), -0.7\% (Sports), and +3.7\% (Baby). Similar trends were observed when replacing modal
embeddings with constant vectors. Metrics fluctuated slightly across datasets and methods, suggesting model performance relies mainly on user item interaction signals rather than multimodal semantics.

To investigate this phenomenon more deeply, we selected the most common multimodal recommendation baseline VBPR for a more granular analysis. By systematically manipulating the numerical distribution of the replaced embeddings, we observed the following behaviors:

\begin{itemize}
  \item \textbf{Degradation to BPR:} When the modality embedding is set to all zeros, VBPR effectively reduces to the standard BPR model. Its performance becomes almost indistinguishable from that of BPR, and as shown in \autoref{fig:sub1}, their performance curves closely align under varying values of the regularization weight (\texttt{reg\_weight}). This phenomenon is fully consistent with theoretical expectations.

\item \textbf{Impact of Constant Value:} In \autoref{fig:sub1}, we observe that BPR and VBPR with large regularization weights lead to model collapse, causing the Recall@10 values to approach nearly zero. In \autoref{fig:sub2}, we find that increasing the constant values can partially alleviate this collapse and restore the model to normal performance.

  \item \textbf{Distribution Dependence:} Under very large regularization, model performance, smaller variance under the best constant of the mean of 1.0, further enhances performance; in our experiments, a standard deviation of 0.0 surpassed 0.4 as described in \autoref{fig:sub3}.
\end{itemize}

In these simple fusion models, an input with no semantic content can outperform a meaningful one if its magnitude is sufficiently large. Adjusting its standard deviation within an appropriate range can further enhance performance. We attribute the root cause of this phenomenon to the following factors:

\begin{itemize}

    \item \textbf{With constant embeddings}, the modality term in VBPR effectively becomes a bias term. The magnitude of this bias is directly controlled by the magnitude of the input constant, explaining why performance increases with the constant's value.

    \item \textbf{With noise embeddings}, the modality term introduces random fluctuations to each item's score, with the variance of these fluctuations proportional to that of the input noise. Within a certain range, a larger variance provides the model with a broader dynamic range for adaptation during training, thereby enhancing its robustness over the learning process.
\end{itemize}

This serves as a critical insight: when evaluating multimodal models, merely demonstrating superior performance over a baseline (e.g., VBPR $>$ BPR) is insufficient proof of efficacy. Such performance gains may originate from architectural artifacts rather than genuine multimodal learning.

\section{Conclusion and Future Work}
This study critically examines whether current multimodal recommendation models truly leverage all available modality information. Our analysis first reveals that most existing multimodal recommendation models primarily rely on multimodal embeddings, particularly those equipped with sophisticated fusion mechanisms. Surprisingly, some classic baselines with simple fusion strategies, such as VBPR and BM3, exhibit unexpectedly stable performance even when their multimodal embeddings are replaced with meaningless vectors. A closer investigation of individual modalities further shows that text plays a dominant role: most models maintain comparable performance even when image embeddings are removed. We also discuss models whose behavior deviates from this general trend. These findings have important implications for the multimodal recommendation community.


We will continue to monitor developments in multimodal recommendation to ensure fair and rigorous comparisons. Moreover, we plan to explore the design of new models informed by our insights into fusion strategies and the pivotal role of text in multimodal learning.

%
%
%
\bibliographystyle{splncs04}
\bibliography{base}
\end{document}